\documentclass[12pt]{iopart}
\expandafter\let\csname equation*\endcsname\relax
\expandafter\let\csname endequation*\endcsname\relax\usepackage{amsmath}
\usepackage{hyperref}
\usepackage{graphicx}
\usepackage{import}
\usepackage{float}
\usepackage{subfigure}
\usepackage{ulem}
\usepackage{cite}
\usepackage{ctable} % for \specialrule command

\begin{document}

\title[]{Charge dependence of neoclassical and turbulent transport of light impurities on MAST}

\author{S. S. Henderson$^{1,2}$, L. Garzotti$^2$, F. J. Casson$^2$, 
        D. Dickinson$^{2,3}$, M. O'Mullane$^1$, A. Patel$^2$, 
	      C. M. Roach$^2$, H. P. Summers$^1$, H. Tanabe$^4$, M. Valovi\v{c}$^2$ 
	      and the MAST team}

\address{$^1$ Department of Physics SUPA, University of Strathclyde, Glasgow, G4 ONG, UK}
\address{$^2$ EURATOM/CCFE Fusion Association, Culham Science Centre, Abingdon, Oxfordshire, OX14 3DB, UK}
\address{$^3$ York Plasma Institute, Department of Physics, University of York, Heslington, York, YO10 5DD, UK}
\address{$^4$ Graduate School of Frontier Sciences, University of Tokyo, Tokyo, 113-8656, Japan}

\ead{stuart.henderson@ccfe.ac.uk}

\begin{abstract}
Carbon and nitrogen impurity transport coefficients are determined from gas puff experiments carried out during repeat L-mode discharges on the Mega-Amp 
Spherical Tokamak (MAST) and compared against a previous analysis of helium impurity transport on MAST. The impurity density profiles are measured on 
the low-field side of the plasma, therefore this paper focuses on light impurities where the impact of poloidal asymmetries on
impurity transport is predicted to be negligible. A weak screening of carbon and nitrogen is found in the plasma core, whereas the helium density 
profile is peaked over the entire plasma radius. Both carbon and nitrogen experience a diffusivity of the order of \mbox{10 m$^2$/s} and a strong inward 
convective velocity of \mbox{$\sim40$ m/s} near the plasma edge, and a region of outward convective velocity at mid-radius. The measured impurity 
transport coefficients are consistent with neoclassical Banana-Plateau predictions within $\rho\leq0.4$. Quasi-linear gyrokinetic predictions of the 
carbon and helium particle flux at two flux surfaces, $\rho=0.6$ and $\rho=0.7$, suggest that trapped electron modes are responsible for the anomalous 
impurity transport observed in the outer regions of the plasma. The model, combining neoclassical transport with quasi-linear turbulence, is shown to 
provide reasonable estimates of the impurity transport coefficients and the impurity charge dependence.
\end{abstract}

\section{Introduction}\label{sec:intro}

Tokamaks contain a range of partially and fully ionised impurity ions. Helium ash will be unavoidable in D-T plasma, while plasma 
facing materials in ITER will consist of tungsten and beryllium. Additionally, seeded elements, such as argon, neon or nitrogen, have been identified as 
suitable mantle and divertor radiators \cite{Kallenback2013}. A build-up of these impurities in the main chamber plasma may 
result in fuel dilution and radiative cooling of the plasma core, degrading the tokamak performance. Predicting and controlling impurity transport in next-step 
tokamaks requires the rigorous testing of suitable models against experimental data and an assessment as to how impurity transport depends on plasma parameters.
Several attempts have already been made in this direction (see Guirlet et al. \cite{Guirlet2006} for example), and this paper focuses particularly on the charge dependence, in the case of a 
spherical tokamak (ST).

The atomic number, $Z$, and mass number, $A\approx m_I/m_p$, both strongly influence neoclassical and turbulent impurity transport. This paper focuses on
fully ionised impurity ions with $Z\leq7$ and $A/Z\sim2$. In this limit, turbulent diffusion can either decrease or increase with Z, depending on the character of the 
dominant turbulent modes \cite{Angioni2006}. The turbulent convective velocity includes contributions from thermodiffusion, rotodiffusion and parallel 
compression \cite{Angioni2006}. The first two terms decrease and increase respectively with $Z$, while parallel compression is proportional to $Z/A$ and is 
therefore approximately constant for fully ionised impurities. It is worth noting however that the magnitude of the turbulent diffusion and convective velocity 
is expected to saturate for fully ionised impurity ions with $Z>10$ \cite{Angioni2006}. In stationary, low collisionality plasma neoclassical diffusivity 
decreases with $Z$ \cite{Wenzel1990}, while in highly rotating plasma neoclassical diffusivity becomes a complex function of $Z$ and $A$ \cite{Wong1987}. 
Neoclassical convective velocity scales linearly with $Z$ in all collisionality regimes \cite{Wenzel1990} and can therefore dominate over 
the equivalent turbulent rates for highly charged ions, whereas turbulent diffusion typically always dominates over neoclassical diffusive rates.   

In light of this complex $Z$ dependence of impurity transport, a large number of experimental studies have focused on measuring low-$Z$ 
($Z<10$) \cite{Synakowski1990,Synakowski1993,Wade1995}, medium-$Z$ ($10\leq Z \leq 18$) \cite{Pasini1990,Dux1999,Giroud2007} and high-$Z$ ($Z>18$) 
\cite{Dux2004,Valisa2011,Casson2015} impurity transport coefficients. The majority of these experiments have been carried out on conventional 
tokamaks with only a limited number of experimental studies on STs \cite{Stutman2003,Delgado2011,Scotti2013,Henderson2014}. 
The magnetic field and ion temperature is typically lower in STs than in conventional tokamaks, and so ST data tests impurity transport models in
plasmas where the neoclassical transport is higher and more dominant. Additionally, different modes can drive turbulent transport and 
long wavelength ion temperature gradient (ITG) modes, often associated with anomalous transport on conventional tokamaks, are frequently 
stabilised in STs by sheared toroidal flows \cite{Roach2005}. It was found in a previous study that helium impurity ions are sensitive to trapped electron mode (TEM) 
turbulence in L-mode discharges in the Mega-Amp Spherical Tokamak (MAST) \cite{Henderson2014}. This paper extends that study using measurements 
of carbon and nitrogen transport made during L-mode MAST plasma scenarios. Nitrogen and carbon are analysed in this paper to assess the transport
of extrinsic and intrinsic impurities respectively, and the $Z$ dependence of impurity transport in these MAST plasmas is obtained by comparing with 
the helium results in \cite{Henderson2014}.

Experimental transport coefficients are determined using the 1.5-D transport code \textsc{sanco} \cite{Lauro-Taroni1994} to reproduce the 
density profile evolution of C$^{6+}$ and N$^{7+}$, measured following a short impurity gas puff using charge exchange (CX). Predictions of the
carbon, nitrogen and helium \cite{Henderson2014} transport coefficients were obtained using the neoclassical code \textsc{neo} \cite{Belli2008,Belli2012} 
and the gyrokinetic code \textsc{gkw} \cite{Peeters2009}. The results suggest that, compared to helium, carbon and nitrogen have higher rates of 
inward convective velocity and diffusion near the plasma edge, and the convective velocity of carbon and nitrogen (but not of helium) changes sign from
inwards to outwards at mid-radius. Carbon, nitrogen and helium diffusivities agree within error bars in the core. Our theory based impurity transport model 
suggests that the trend in $Z$ near the edge is caused by TEM turbulence, and finds that the reversal in the convective velocity for carbon and nitrogen
at mid-radius is consistent with neoclassical transport in the Banana-Plateau regime.

This paper is organised as follows. A description of the L-mode 900 kA plasma used for the analysis is given in section \ref{sec:2}, where we also 
provide details of the diagnostics used to measure the CX spectral intensities and the bulk plasma parameters that are required as inputs to 
the simulation codes. In section \ref{sec:3}, we present the model of the C$^{6+}$ and N$^{7+}$ density evolution, and describe the procedure used to 
obtain the experimental transport coefficients. We discuss the measured impurity transport coefficients and their dependence on $Z$ in section \ref{sec:4}, 
and interpret these results in terms of an impurity transport model based on neoclassical and quasi-linear gyrokinetic theory. Lastly, we provide a 
brief conclusion in section \ref{sec:5}.

\section{Plasma Scenario and Diagnostics}\label{sec:2}

\begin{table}[bp]
\caption{Details of the injected gas species during repeated MAST discharges.} % title of Table
\centering % used for centering table
\begin{tabular}{ccc} % centered columns (4 columns)
\specialrule{.1em}{.05em}{0.05em} 
Mast Shot \#   & Injected Gas Species   & Fuelling Duration / ms \\ [0.5ex] % inserts table heading
\hline % inserts single horizontal line
  28052          & None                   & ---                \\   
  28053          & CH$_4$                 & 35                 \\   
  29261          & He                     & 15                 \\   
  29427          & None                   & ---                \\   
  30439          & N$_2$                  & 15                 \\   
\specialrule{.1em}{.05em}{0.05em} 
\end{tabular}
\label{table:mastgas} % is used to refer this table in the text
\end{table}

We have chosen to study the dependence on $Z$ of impurity transport in the L-mode 900 kA MAST discharges described in our initial report \cite{Henderson2014}. 
This plasma is free from edge localised modes (ELM), while TEM turbulence is expected in the outer region of the plasma. The transport analysis time 
window is reduced to a few tens of milliseconds during the current flat-top (beginning at $t\sim0.2$ s) to avoid the magnetohydrodynamic mode (MHD) activity occurring at $t\sim0.28$ s.
A summary of the key plasma parameters are as follows: plasma current $I_p=900$ kA, toroidal magnetic field $B_T=0.55$ T, NBI heating power $P_{NBI}=2.1$ MW, on-axis electron density 
$n_e=3.5\times10^{19}$ m$^{-3}$, on-axis electron temperature $T_e=1.0$ keV, $I_p$ flat-top and diffusion duration is of the order of 100 ms, 
global energy confinement time is of the order of 10 ms, major radius $R_0=0.83$ m, minor radius $a=0.6$ m, elongation $\kappa=1.93$ and triangularity 
$\delta=0.4$. Plasma parameters vary by less than 20 \% during the $I_p$ flat-top in this particular discharge. Time traces of various equilibrium quantities are shown in \cite{Henderson2014}.

Helium (He), methane (CH$_4$) and nitrogen (N$_2$) gas were each injected into the plasma at the beginning of the $I_p$ flat-top using a piezo-valve located on the inboard side 
of the lower divertor. The transport analysis time window is taken during the $I_p$ flat-top time between $0.21$ s $\leq t_{avg} \leq 0.27$ s, which is free from the long-lived mode and sawtooth 
MHD activity. The plasma scenario was repeated three times to study one impurity gas puff (of the order of 10 ms) per pulse. Two further repeat pulses were carried out with no 
impurity gas injection for comparison. A summary of the shot numbers and injected impurity gas is given in table \ref{table:mastgas}. A longer fuelling time was needed for CH$_4$ 
because the measured fuelling efficiency for carbon was found to be lower than that for helium \cite{HendersonThesis}. The impurity injections can be considered trace since they 
increased the effective charge of the plasma, $Z_{eff}$, by $\leq5$ \%.

\begin{figure}[t]
	\centering
		\includegraphics[width=0.7\textwidth]{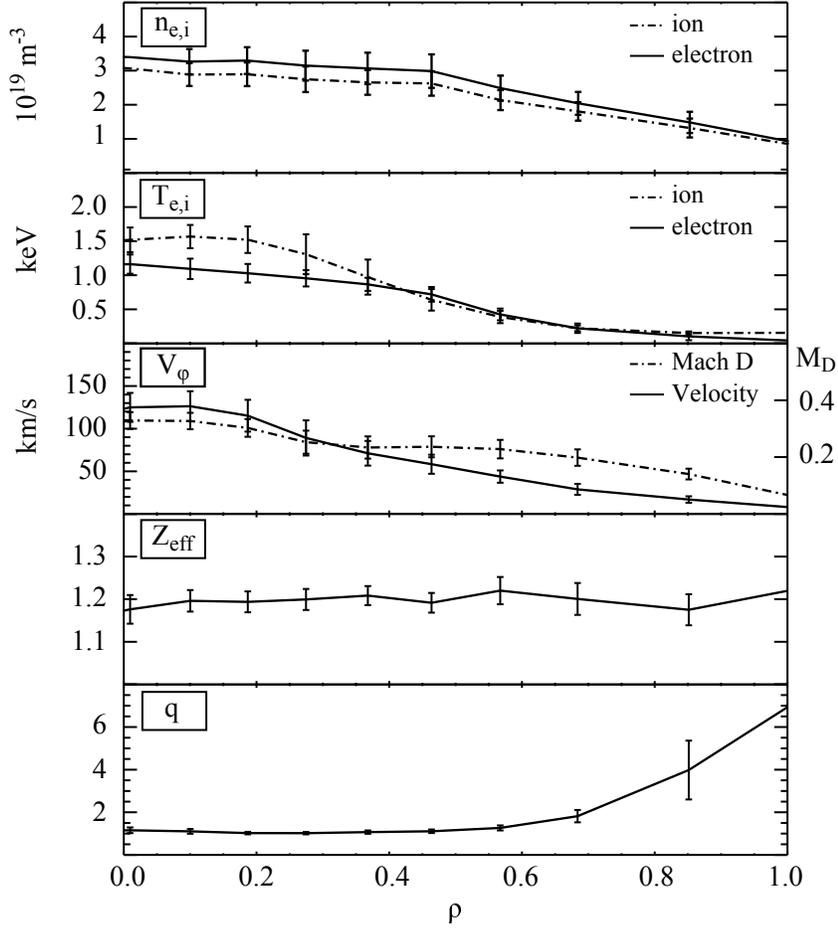}
	\caption{Profiles of (from top to bottom) the electron and ion density, electron and ion temperature, ion toroidal velocity and (deuterium) Mach number, effective 
	charge, and the safety factor, as a function of the normalised toroidal magnetic flux, $\rho=\sqrt{\phi_N}$. Each radial point represents the weighted mean over 
	the five repeated shots over the time interval $0.21$ s $\leq t_{avg} \leq 0.27$ s, with the error bar indicating the standard deviation. Both the ion density and 
	effective charge have been calculated using CX concentrations of carbon ($\sim0.1$ \%) and helium ($\sim1$ \%). }
	\label{fig:2.1}
\end{figure}

The transport analysis presented in section \ref{sec:4} is supported by simulation codes which require, as input, a description of the bulk plasma equilibrium. 
MAST is equipped with a comprehensive set of diagnostics. The Thomson scattering system measures the electron temperature $T_e$ and density $n_e$ profiles 
with a radial resolution and sampling rate of 10 mm and 240 Hz respectively \cite{Scannell2008}. Spectrally resolved CX measurements of C VI from the 
CELESTE-3 diagnostic are used to deduce the ion temperature $T_i$ and toroidal rotation $V_{\phi}$ profiles with radial and temporal 
resolutions of 1 cm and 4 ms \cite{Conway2006}. Radial profiles of the He$^{2+}$ and C$^{6+}$ concentration (typically of the order of $\sim1$ \% and $\sim0.1$ \% respectively) 
are derived from calibrated CX measurements from the RGB filtered imaging diagnostic \cite{Patel2004} and used to determine self-consistent profiles of $Z_{eff}$ and
the main ion density $n_i$ \cite{HendersonThesis}. Radial and temporal resolutions of 1 mm and 5 ms respectively are available from RGB, although typically the radial resolution is
reduced to $1-5$ cm in order to improve statistics. The magnetic equilibrium, and hence safety factor $q$, has been reconstructed using the \textsc{efit++} code \cite{Appel2006}, 
with $D_{\alpha}$ and motional Stark effect (MSE) constraints \cite{DeBock2008}. 

Radial profiles of several of the plasma parameters described above are shown in figure \ref{fig:2.1} as a function of $\rho=\sqrt{\phi_N}$, where $\phi_N$ is the normalised toroidal 
magnetic flux. Each radial point represents the weighted mean, with a corresponding standard deviation, taken over time ($t_{avg}=0.21-0.27$ s with 5 ms 
resolution), shot (5 repeat discharges) and space. Significant variations in the transport coefficients are not evident for spatial resolutions of $<5$ cm, therefore the 
plasma parameters are spatially resolved to 5 cm. Uncertainties of $\leq20$ \% and $\leq40$ \% are found for the weighted mean values 
and their radial gradients respectively. The standard deviation of the radial gradient was calculated using the method of propagation of errors.

Measurements of the active CX C VI $n=8\rightarrow7$ ($\lambda=529.1$ nm) and N VII $n=9\rightarrow8$ ($\lambda=566.9$ nm) spectral line 
intensities, induced by NBI, have been made with a spectrometer and multi-chord setup. For carbon we use the existing CELESTE-3 diagnostic \cite{Conway2006} 
(primarily measuring $T_i$ and $V_{\phi}$) consisting of 64 active and passive tangential fibres with a spatial resolution of $\sim1$ cm. A similar setup is used
for nitrogen but with 16 active and passive fibres with a spatial resolution of $\sim5$ cm. Integration times of 5 ms are used for both measurements. 

\section{Impurity Density Evolution}\label{sec:3}

\begin{figure}[t]
       \centering
	       \includegraphics[width=0.6\textwidth]{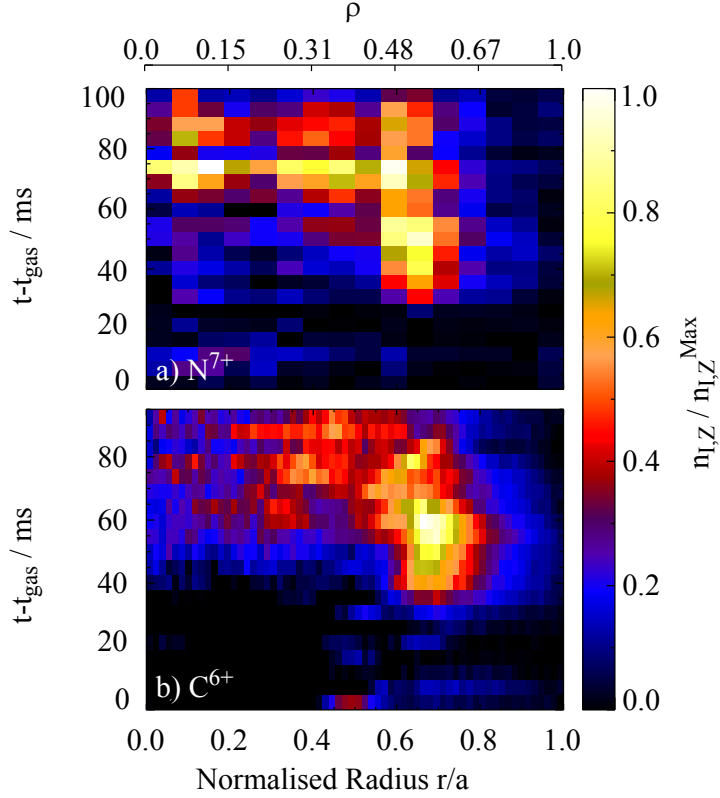}
       \caption{Profile evolution of the (a) N$^{7+}$ and (b) C$^{6+}$ impurity density profile normalised to the maximum value. 
       The start time of the gas puff is subtracted from the time axis in both graphs. Both radial flux labels $\rho$ and $r/a$ 
       are indicated above and below the plots respectively. $t_{gas}$ corresponds to (a) 195 ms and (b) 175 ms.  }
       \label{fig:3.1}
\end{figure}

The C VI and N VII CX spectral line intensities are converted into C$^{6+}$ and N$^{7+}$ density profiles using the CX effective emission coefficients interpolated
from the Atomic Data Analysis Structure (ADAS) \cite{Summers2004} and simulations of the line-integrated neutral beam density \cite{HendersonThesis}. For the impurity transport analysis, 
only the evolution of the impurity density profile shape matters as explained later in this section; absolute concentrations are not strictly necessary. Therefore the 
systematic error from the thermal beam halo, which can increase the CX intensity by a constant radial fraction of $\sim20$ \% for the same impurity density \cite{HendersonThesis}, will not modify 
the derived transport coefficients. Secondary impurity `plume' emission, induced by electron collisional excitation of the partially ionised impurity `plume' ions left 
over following the CX process, can alter the radial CX intensity profile shape. However, we expect any changes in the profile shape caused by impurity plume emission 
to be negligible due to the relatively long ionisation lengths and low electron excitation rate coefficients associated with the plume ions \cite{HendersonThesis}. 
The C$^{6+}$ and N$^{7+}$ density evolution after the gas puff is illustrated in figure \ref{fig:3.1}. Intrinsic impurity density profiles, calculated during the two 
reference discharges, have been subtracted. A region of $\sim20$ ms where no signal from the gas puff is measured is found immediately after the piezo-valve is triggered. 
This corresponds to the finite time taken by the impurity gas to travel through the plenum and into the main chamber. 

\begin{figure}[t]
	\centering
		\includegraphics[width=0.6\textwidth]{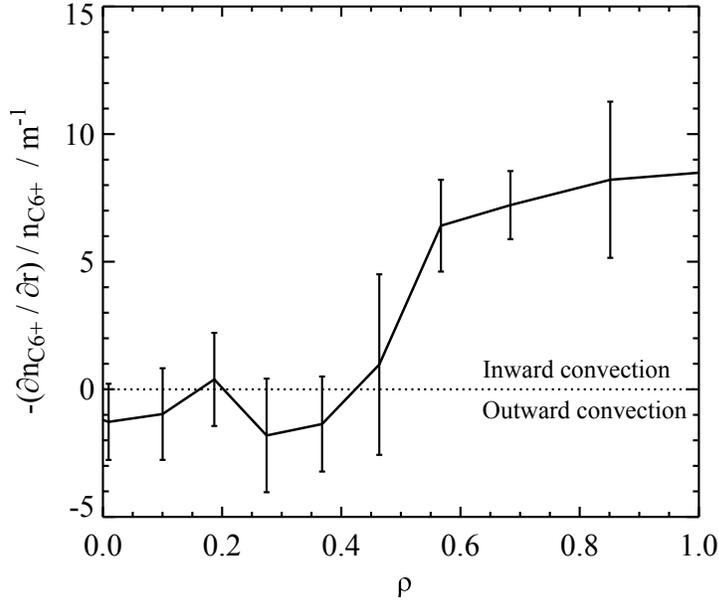}
	\caption{A profile of the intrinsic logarithmic C$^{6+}$ density gradient from the discharges without impurity gas puffs. Each radial point represents the weighted mean 
	over time ($t_{avg}=0.21-0.27$ s), space and shot (2 repeat discharges with no impurity gas puff), with the error bars denoting a corresponding standard deviation.}
	\label{fig:3.2}
\end{figure}

The 1.5-D radial transport code \textsc{sanco} \cite{Lauro-Taroni1994} is used to numerically solve the continuity equation for the density of each impurity ionisation 
stage with a given set of boundary conditions. \textsc{sanco} uses a cylindrical geometry, therefore the transport equations are written as
\begin{eqnarray}\label{transport_equations}
\frac{\partial n_{I,Z}}{\partial t}&=& -\frac{1}{r}\frac{\partial}{\partial r}\left( r\Gamma_{I,Z}\right)+S_{I,Z} \\
\Gamma_{I,Z}&=&-D_I\frac{\partial n_{I,Z}}{\partial r}+V_I n_{I,Z} 
\end{eqnarray}
where $D_I$ and $V_I$ are the diffusion and convective velocity coefficients for impurity species $I$ and $r=\sqrt{V_{flux}/2\pi^2R_0}$ is the effective radius of the 
flux surface enclosing a volume, $V_{flux}$. Ionisation and recombination rate coefficients provided by ADAS are used to determine the source term, $S_{I,Z}$. 
Determining the transport coefficients requires a least-squares minimisation of the error between the experimental and modelled fully ionised impurity density evolution by 
the appropriate optimisation of $D_I$ and $V_I/D_I$. 

The \textsc{sanco} model assumes that the impurity transport coefficients are constant in time and equal for each impurity ionisation stage. The first assumption is 
valid for the time window of $t_{avg}$ because the plasma is approximately in steady-state during the plasma current flat-top on MAST. In theory, the latter assumption 
is not strictly valid, since our main hypothesis is that there \textit{is} a charge dependence on impurity transport; however this simplification is valid in this 
case because the fully ionised impurity ions are dominant over the majority of the plasma radius. Ionisation timescales for the partially ionised impurity ions that 
exist near the cooler plasma boundary are such that any differences in plasma transport between each ionisation stage will have a negligible effect on the derived 
impurity transport coefficients.

\begin{figure}[t]
	\centering
		\includegraphics[width=0.95\textwidth]{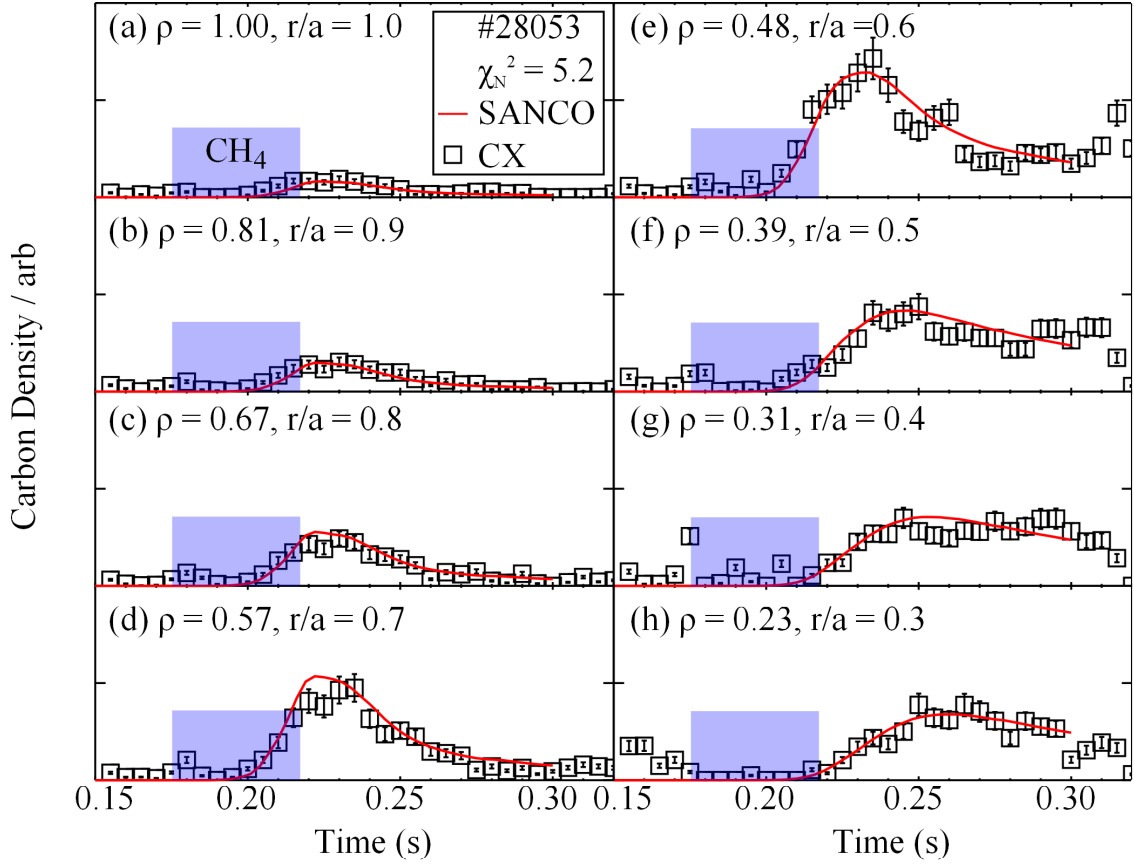}
	\caption{A comparison between the temporal evolution of the experimental and simulated \textsc{sanco} C$^{6+}$ density profile for a number of different radii. The blue shaded
	region indicates the timing of the impurity gas puff.}
	\label{fig:3.3}
\end{figure}

Radial impurity density profiles from \textsc{sanco} are averaged over flux surfaces and hence do not account for any poloidal variation. Low-field side (LFS) localisation of 
impurities is an important issue for transport studies on STs because a low aspect ratio and strong toroidal rotation, typical of ST plasma, can enhance 
the poloidal asymmetry that arises from the outward centrifugal force \cite{Wesson1997}. Unfortunately, CX impurity line intensities fall below the detection threshold of the diagnostics on the high-field side (HFS) 
of the plasma due to the exponential fall-off in neutral beam density, therefore the poloidal variation of the impurity density cannot be measured. However, a prediction 
of the LFS localisation can be made using \cite{Wesson1997}
\begin{equation}\label{LFSlocalisation}
\frac{n_{I,Z}^{LFS}}{<n_{I,Z}>}=2\epsilon\frac{m_Z}{m_i}M_i^2\left(1-\frac{Zm_i}{m_Z}\frac{Z_{eff}T_e}{T_e+T_i}\right)
\end{equation}
where $n_{I,Z}^{LFS}$ is the impurity density at the low-field side of the plasma, $\epsilon$ is the inverse aspect ratio and $<n_{I,Z}>$ is the average impurity 
density on a flux surface. With profiles of $M_i$, $Z_{eff}$, $T_e$ and $T_i$ from figure \ref{fig:2.1}, values of $n_{C6}^{LFS}/<n_{C6}>\leq0.1$ are found 
and therefore any change in the derived carbon transport coefficients is expected to be within the fitting error. The poloidal variation of heavier impurities in 
similar plasma conditions may be significant; for example, a trace injection of argon with dominant ionisation state Ar$^{16+}$ could have values of $n_{Ar16}^{LFS}/<n_{Ar16}>\sim0.4$. 

To fit the transport coefficients, first we parameterise $D_I$ and $V_I/D_I$ over 10 equally spaced radial points corresponding to a spatial resolution of 5 cm. 
Fixed values of $V_I/D_I$ are used for the first iteration of the minimisation with $D_I$ as a free parameter. Measurements of the C$^{6+}$ logarithmic radial density 
gradient made during the two reference discharges provide an estimate of the zero-flux ($\Gamma_{I,Z}=0$) peaking factor and hence an initial guess for $V_I/D_I$: 
\begin{eqnarray}\label{peaking_factor}
V_I/D_I=-\frac{1}{L_{n_{I,Z}}}=\frac{\partial \ln{n_{I,Z}}}{\partial r}
\end{eqnarray}
The measured zero-flux peaking factor profile shown in figure \ref{fig:3.2} suggests an inward pinch in the region $0.5\leq\rho\leq1.0$ and a region of outward convective 
velocity between $0.3\leq\rho<0.5$. Next we set both $D_I$ and $V_I/D_I$ as free parameters allowing for any correction to $V_I/D_I$. After $D_I$ and $V_I$ have been determined
over the equally spaced radial grid, we then remove grid points within the regions of approximately constant $D_I$ and $V_I$. The errors, illustrated by the shaded region 
in figure \ref{fig:3.4}, represent one standard deviation for $D_I$ at constant $V_I/D_I$ and for $V_I/D_I$ at constant $D_I$. 

The edge model in \textsc{sanco} requires inputs for the neutral influx $\Gamma^{gas}$, the recycling coefficient $R_{rcy}$, the parallel loss time $\tau_{||}$, the neutral kinetic energy 
$E_{th}$, the fuelling efficiency $\epsilon_F$ and the values of $D_I^{SOL}$ and $V_I^{SOL}$ in the scrape-off-layer (SOL) \cite{Giroud2007}. $\Gamma^{gas}$ is calculated by integrating the 
measured impurity density over the plasma volume. We set $R_{rcy}=0.01$ to represent the low recycling typical of carbon and $\tau_{||}=1$ ms which was based on the SOL connection 
length and $E_{th}$. The value of $E_{th}$ determines the distance that the neutral particles travel into the plasma before ionising: e.g. with $E_{th}=2$ eV the neutrals travel $\sim10$ cm into 
the plasma (corresponding to region between $0.8\leq\rho\leq1.0$). Underestimating $E_{th}$ will cause $D_I$ and $V_I$ to be overestimated over the range determined by the maximum penetration depth of the neutrals. 

Atoms and molecules leaving the gas valve have thermal energies corresponding to room temperature. As they interact with the scrape-off-layer (SOL) plasma, the molecules dissociate at a threshold $T_e$ forming atoms 
with Franck-Condon breakup energies of a few eV. The atoms are kinetically energised by elastic collisions and CX reactions with plasma ions, however their ionisation 
times are too short for them to thermalize fully with the SOL plasma (generally around 10 - 20 eV on MAST) and therefore we assume that $E_{th}\sim2$ eV 
\cite{Tamor1981,Giroud2007}. We note that increasing $E_{th}$ to around 10 eV caused a reduction of $V_I$ and $D_I$ within the region $0.7\leq\rho\leq1.0$, but this value of
$E_{th}$ is much higher than that in MAST. The values of $\epsilon_F$, $D_I^{SOL}$ and $V_I^{SOL}$ were tailored to match the magnitude of the \textsc{sanco} and 
CX impurity density profiles up to the expected penetration depth of the neutral impurity atoms ($0.8<\rho\leq 1.0$).

A comparison of the \textsc{sanco} and experimental impurity density evolution is shown in figure \ref{fig:3.3}. The impurity particle flux perturbation induced by the gas puff
was not evident within the plasma core $\rho\leq0.2$, therefore the transport analysis is restricted to the spatial range of $0.2\leq\rho\leq0.8$. As a goodness of fit metric, we 
use the reduced chi-squared statistic, $\chi^2_{red}=\sum_i{(f_i-y_i)^2}/\sigma^2/DOF$ where $i$ is the number of measurements, $\sigma$ is the variance and $DOF$ is the degrees of freedom. The C$^{6+}$ density 
fit (between $0.21<t<0.27$ s) produced a $\chi^2_{red}=5.2$, while the N$^{7+}$ density fit was quantified by $\chi^2_{red}=4.4$. A model in accordance with the variance should give 
$\chi^2_{red}=1$, while $\chi^2_{red}>>1$ indicates that the model is incorrect and $\chi^2_{red}>1$ indicates that $\sigma$ has been underestimated. For both C$^{6+}$ and N$^{7+}$ we set $\sigma=0.2$ 
however, assuming $DOF\approx100$, a $\chi^2_{red}=5.2$ would suggest a more realistic (but still modest) variance of $\sim2\sigma$. 

\section{Charge Dependence}\label{sec:4}

\begin{figure}[t]
	\centering
		\includegraphics[width=0.7\textwidth]{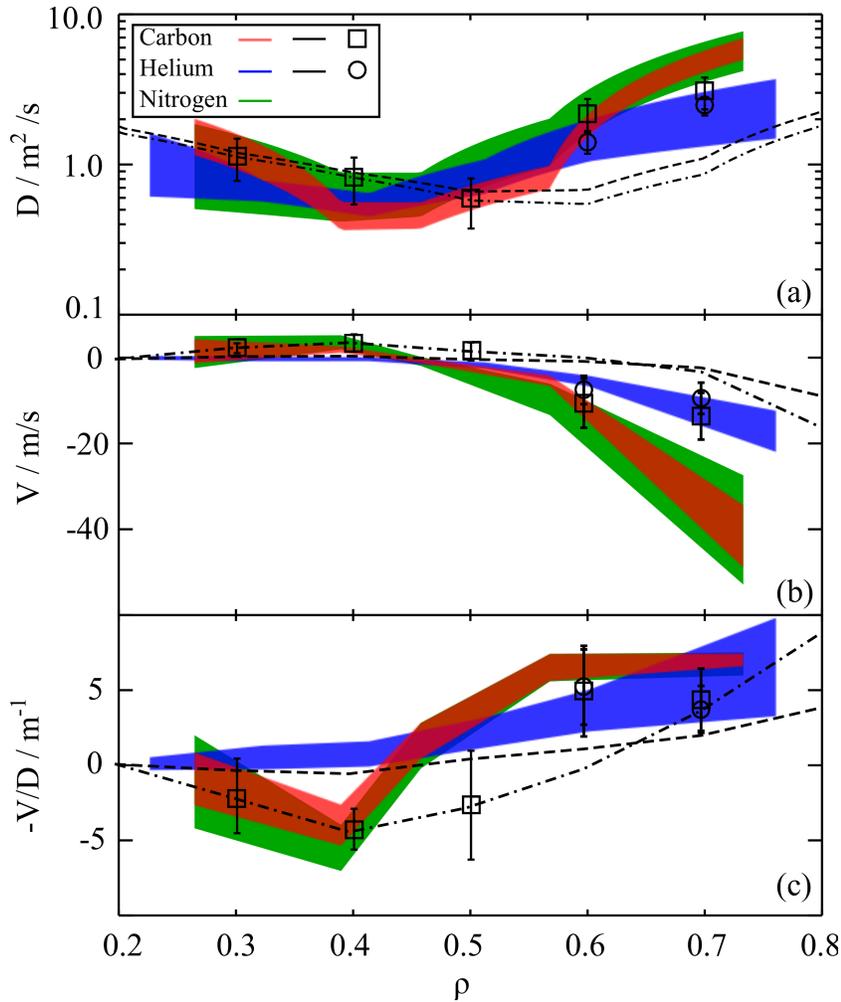}
	\caption{Radial profiles of impurity transport coefficients for helium, carbon and nitrogen: (a) the diffusivity, $D$; (b) the convective velocity, $V$; and (c) the 
	         corresponding steady state impurity peaking factor, $-V/D$. \textsc{neo} simulations of the neoclassical impurity transport coefficients for helium and carbon 
	         are shown as dashed and dashed-dotted lines. The quasi-linear and neoclassical transport coefficients for helium and carbon have been added together at 
		 $\rho=0.6$ and $\rho=0.7$ as shown by the symbols. The symbols within $\rho\leq0.5$ indicate the typical neoclassical error.}
	\label{fig:3.4}
\end{figure}

%======================
% Experimental trends
%======================

The $D_I$, $V_I$ and $V_I/D_I$ \textsc{sanco} profiles are shown in figure \ref{fig:3.4}. Carbon and nitrogen transport coefficients agree within error bars over the 
analysed radial region. Both experience a strong inward pinch ($0.5\leq\rho\leq0.8$), a region of weak outward convective velocity around mid-radius ($0.4\leq\rho<0.5$), and a
diffusive region in the plasma core ($\rho<0.4$). Compared to helium, they experience a larger inward pinch and diffusivity near the plasma edge and a change of
convective velocity direction (from inwards to outwards) at mid-radius. This suggests that higher $Z$ impurities in an ST will stagnate around the plasma mid-radius, whereas
lower $Z$ impurities will accumulate further into the plasma core.

%======================
% Describe the models
%======================

We model the neoclassical particle flux at each radial point using the local drift kinetic neoclassical code \textsc{neo} \cite{Belli2008,Belli2012}. 
Toroidal rotation of the bulk plasma and self-consistent concentrations of C$^{6+}$ and He$^{2+}$ from CX measurements are included in the 
\textsc{neo} inputs. Neoclassical predictions for helium and carbon are shown by the dashed and dashed-dot lines respectively in figure \ref{fig:3.4}; rates for nitrogen differ little from carbon and are therefore omitted. 
In the region $0.4<\rho\leq0.7$, neoclassical convective velocity and diffusion are lower in magnitude than experiment for both impurities. Carbon and helium diffusivities 
agree with the neoclassical predictions in the region $\rho\leq0.4$. Neoclassical simulations reproduce the reversal of the sign of the impurity convective velocity with radius,
which is observed for carbon but not for helium. At $\rho\sim0.4$ both C$^{6+}$ and He$^{2+}$ ions are in the low collisionality Banana-Plateau (BP) regime and hence, assuming that $T_i=T_I$, the 
neoclassical BP convective velocity is (see Eq. 3 of \cite{Wenzel1990})
\begin{equation}\label{banana_convective velocity}
V^{bp}_I=D^{bp}_I Z\left[\frac{\partial \ln{n_i}}{\partial r}+\left(\frac{3}{2}\left[\frac{1}{Z}-1\right]-\frac{1}{Z}\frac{k_{1,2}^I}{k_{1,1}^I}+\frac{k_{1,2}^i}{k_{1,1}^i} \right)\frac{\partial \ln{T_i}}{\partial r} \right]
\end{equation}
where $D^{bp}$ is the BP diffusivity and $k_{i,j}^a$ are the viscosity coefficients. The direction of the convective velocity is governed by the density and temperature logarithmic gradients, 
and by the factor in brackets multiplying the temperature gradient. To illustrate the dependence on $Z$ of this multiplying factor, we use the asymptotic form of $k_{i,j}^a$ (defined in Eq. A.15 
of \cite{Wenzel1990}) with the dominant contribution from the banana regime. In this limit, the multiplying term reduces to $\sim-0.19$ and $\sim-0.34$ for helium and carbon respectively.
Since the temperature gradient is also negative, increasing $Z$ therefore enhances the screening of the impurity ions providing an explanation for the reversal in the impurity convective velocity 
direction observed for carbon.

\begin{figure}[t]
	\centering
		\includegraphics[width=0.95\textwidth]{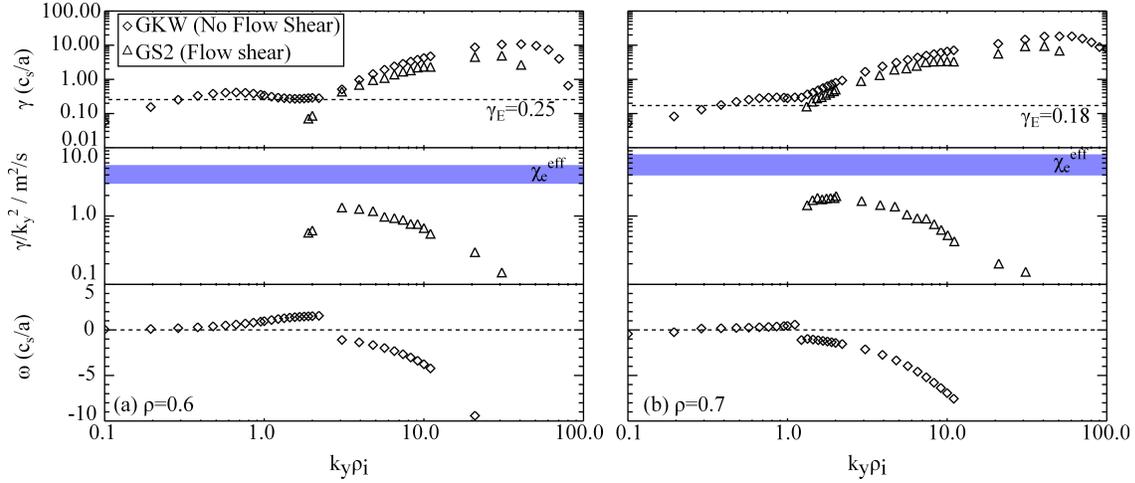}
		
	\caption{Linear growth rate spectra (top panels), mixing length diffusivity estimates (middle panels) and mode frequencies (bottom panels) are shown for (a) $\rho=0.6$ 
	         and (b) $\rho=0.7$. Calculations were performed with and without flow shear using the \textsc{gs2} (triangles) and \textsc{gkw} (diamonds) codes
		 respectively. The dashed lines in the top panels represent the value of the flow shear (normalised to $c_s/a$), while the shaded region in the middle panels represents the effective 
		 electron heat diffusivity calculated using \textsc{jetto} \cite{Cenacchi1988,Cenacchi1988a}.}
	\label{fig:3.45}
\end{figure}

The turbulent particle flux is modelled at $\rho=[0.6,0.7]$ with the local flux-tube gyrokinetic code \textsc{gkw} \cite{Peeters2009} 
\footnote{Input files used for the \textsc{neo} and \textsc{gkw} simulations can be found at \url{https://bitbucket.org/gkw/publication-input/src/HEAD/2015\_Henderson\_PPCF/}}. These
two flux surfaces are in the steep density gradient region where TEMs are linearly unstable \cite{Henderson2014}. Turbulent contributions within $\rho<0.5$ have not been modelled 
since the flatter density gradient stabilises the TEMs. A pure plasma ($n_e=n_i$) has been assumed for the \textsc{gkw} simulations since the low impurity concentrations ($Z_{eff}<1.3$) have negligible effects 
on the quasi-linear particle fluxes \cite{Estrada-Mila2005}. Linear growth rate $\gamma$ spectra calculated with \textsc{gkw} for $\rho=[0.6,0.7]$ 
are shown in the top panels of figure \ref{fig:3.45} as a function of $k_y\rho_i$, where $k_y$ is the binormal perpendicular wavenumber and $\rho_i$ is the
local Larmor radius of the main ion. The stabilising effect of equilibrium toroidal flow shear $\gamma_E=u'B_{\theta}/B\sim u'\epsilon/q$, where $u'$ is 
the radial gradient of the equilibrium toroidal rotation has been determined with the gyrokinetic code \textsc{gs2} \cite{Kotschenreuther1995,Dorland2000} using the technique described 
in \cite{Henderson2014,Roach2009}. The bottom two panels of figure \ref{fig:3.45} show the mixing length diffusivity estimate $\gamma/k_y^2$ and the mode frequency $\omega$. 
The dominant modes, corresponding to the wavenumber of the maximum value of $\gamma/k_y^2$, are within the TEM region; $k_y\rho_i\sim3.0$ and 
$k_y\rho_i\sim2.0$ for $\rho=0.6$ and $\rho=0.7$ respectively. We choose the value of $k_y\rho_i=2.6$ for both flux-surfaces to use in our quasi-linear analysis described below.   

The modelled particle flux can be written as a linear summation of the neoclassical and turbulent contributions,
\begin{eqnarray}\label{theory_particle_flux}
\frac{\Gamma_{I,Z}^{\textsc{model}}}{n_{I,Z}}=\frac{D_I^{\textsc{neo}}}{L_{n_{I,Z}}}+\frac{D_I^{turb}}{L_{n_{I,Z}}}+V_I^{\textsc{neo}}+V_I^{turb}
\end{eqnarray}
where $1/L_{\alpha}=-\partial \ln{\alpha}/\partial r$. The diffusive and convective terms are extracted from the simulation output files 
by including trace species with different gradients in the input files. In the quasi-linear approach we model the saturated turbulence as the dominant linear 
mode with an amplitude that reproduces the effective electron heat diffusivity, $\chi_{e}^{eff}$, calculated using the interpretative transport code 
\textsc{jetto} \cite{Cenacchi1988,Cenacchi1988a} (see table \ref{table:transport}). This allows us to determine the turbulent impurity particle diffusivity
and pinch from their corresponding \textsc{gkw} values, $D^{\textsc{gkw}}$ and $V^{\textsc{gkw}}$, and electron heat diffusivity, $\chi_e^{\textsc{gkw}}$, in linear 
\textsc{gkw} simulations as:
\begin{equation}\label{DnVturb}
D_I^{turb}=\frac{\chi_e^{eff}}{\chi_e^{\textsc{gkw}}}D^{\textsc{gkw}}\mbox{\;,\;\;}V_I^{turb}=\frac{\chi_e^{eff}}{\chi_e^{\textsc{gkw}}}V^{\textsc{gkw}}
\end{equation}

At steady state (i.e. with $\Gamma_{I,Z}=0$) Eq. \ref{theory_particle_flux} provides the modelled dimensionless peaking factor:
\begin{equation}\label{theory_peaking_factor}
\frac{a}{L_{n_{I,Z}}}=-a\frac{V_{I}^{\textsc{neo}}+V_{I}^{turb}}{D_{I}^{\textsc{neo}}+ D_{I}^{turb}}
\end{equation}
where we choose the value of $D_I^{\textsc{gkw}}$, $V_I^{\textsc{gkw}}$ and $\chi_{e}^{\textsc{gkw}}$ associated with the dominant linear mode. Strictly, 
the quasi-linear single value of each of these profiles must be associated with the same $k_y\rho_i$. In most conventional tokamak discharges, 
the ITG modes responsible for impurity particle transport also drive the anomalous heat diffusivity and hence the above method is valid and used in many 
studies \cite{Angioni2011,Casson2013,Angioni2014a}. However on MAST it is the electron turbulence that is thought to drive the anomalous heat transport in most 
discharges \cite{Roach2009}. We therefore assess the applicability of this quasi-linear approach based on its agreement with experiment; a rigorous test would
include a comparison with non-linear gyrokinetic simulations which are not attempted here.

\begin{figure}[t]
	\centering
		\includegraphics[width=0.95\textwidth]{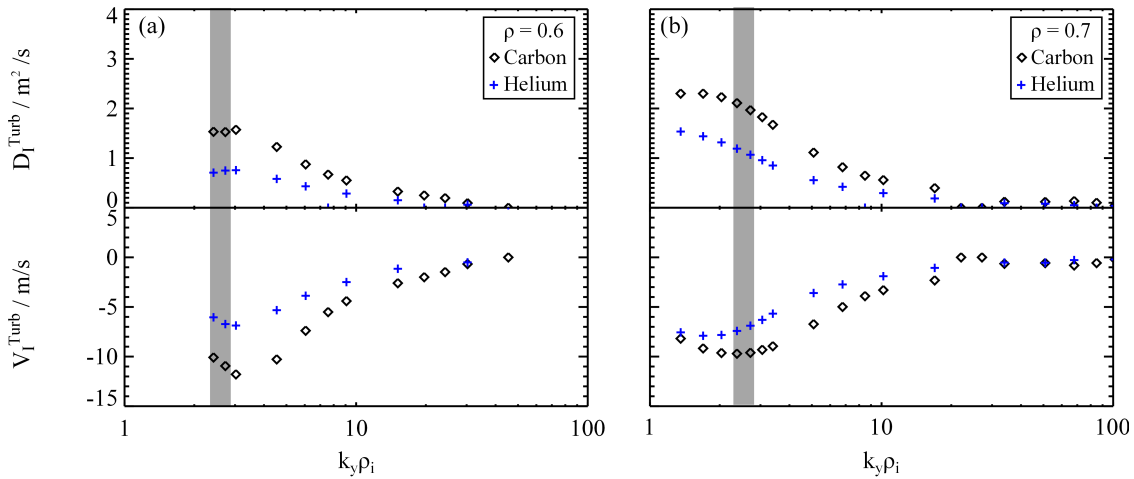}		
	\caption{Quasi-linear carbon and helium transport coefficients from \textsc{gkw} as a function of normalised wavenumber for two flux surfaces, $\rho=0.6$ and $\rho=0.7$. The shaded
	region represents the dominant linear mode numbers for each surface. }
	\label{fig:3.5}
\end{figure}

\begin{table}[bp]
\caption{Comparison of the experimental and modelled transport coefficients for carbon and helium at $\rho=[0.6,0.7]$. The individual contributions to the model from \textsc{neo} and turbulence are
given in the last two columns of the table.} % title of Table
\centering % used for centering table
\begin{tabular}{ccccc} % centered columns (4 columns)
\specialrule{.1em}{.05em}{0.05em} 
			      & Experiment     & Model      \\ [0.5ex] % inserts table heading
			      &                & Total  & \textsc{neo} & Turb    \\ [0.5ex] % inserts table heading
 \underline{$\rho=0.6$}       &		       &		  &		 &		   \\ [0.5ex] % inserts table heading
  $\chi_e$ / m$^2$/s	      &  4.4 $\pm$ 1.3 & ---		  & --- 	 & ---		     \\   
  $D_C$ / m$^2$/s	      &  1.8 $\pm$ 0.3 & 2.1 $\pm$ 0.5    & 0.6          & 1.5  \\   
  $D_{He}$ / m$^2$/s	      &  1.5 $\pm$ 0.5 & 1.4 $\pm$ 0.2    & 0.7          & 0.7  \\   
  $D_C/D_{He}$  	      &  1.2 $\pm$ 0.4 & 1.5              & 0.8          & 2.1  \\   
  $V_C$ / m/s		      &-12.5 $\pm$ 2.4 &-10.5 $\pm$ 5.7   & 0.1          &-10.6 \\   
  $V_{He}$ / m/s	      & -5.5 $\pm$ 1.2 &-7.4 $\pm$ 3.3    &-1.0          &-6.4  \\   
  $V_C/V_{He}$  	      &  2.3 $\pm$ 0.7 & 1.4              &-0.1          & 1.6  \\   
  $-aV_C/D_C$		      &  4.0 $\pm$ 0.5 & 3.0 $\pm$ 1.8    &-0.1          & 4.2  \\   
  $-aV_{He}/D_{He}$	      &  2.2 $\pm$ 0.7 & 3.2 $\pm$ 1.5    & 0.9          & 5.3  \\  
                              &                &                  &              &      \\
 \underline{$\rho=0.7$}       &		       &		  &  	         &      \\ [0.5ex] % inserts table heading
  $\chi_e$ / m$^2$/s	      &  6.1 $\pm$ 2.1 & ---		  & --- 	 & ---	\\   
  $D_C$ / m$^2$/s	      &  4.9 $\pm$ 0.9 & 3.1 $\pm$ 0.7    & 1.0          & 2.0  \\   
  $D_{He}$ / m$^2$/s	      &  2.2 $\pm$ 1.0 & 2.5 $\pm$ 0.4    & 1.4          & 1.1  \\   
  $D_C/D_{He}$  	      &  2.3 $\pm$ 1.1 & 1.2              & 0.8          & 1.8  \\   
  $V_C$ / m/s		      &-34.2 $\pm$ 9.7 &-13.5 $\pm$ 5.4   &-3.9          &-9.6  \\ 
  $V_{He}$ / m/s	      &-12.7 $\pm$ 4.9 & -9.3 $\pm$ 3.6   &-2.3          &-7.1  \\   
  $V_C/V_{He}$  	      &  2.7 $\pm$ 1.3 & 1.4              & 1.7          & 1.4  \\   
  $-aV_C/D_C$	              &  4.1 $\pm$ 0.3 & 2.7 $\pm$ 1.2    & 2.3          & 2.9  \\   
  $-aV_{He}/D_{He}$	      &  3.3 $\pm$ 1.4 & 2.3 $\pm$ 0.9    & 1.0          & 3.8  \\   
\specialrule{.1em}{.05em}{0.05em} 
\end{tabular}
\label{table:transport} % is used to refer this table in the text
\end{table}

Figure \ref{fig:3.5} shows the wavenumber profiles of quasi-linear $D_I^{turb}$ and $V_I^{turb}$ for radial locations $\rho=0.6$ and $\rho=0.7$. 
The impurity transport coefficients peak around $k_y\rho_I\sim1$ (noting that $k_y\rho_C\sim k_y\rho_i/\sqrt{2}$ and $k_y\rho_{He}\sim k_y\rho_i/\sqrt{6}$).   
Turbulent diffusion is moderately higher for carbon compared to helium; a trend previously found using a reduced fluid model for TEMs \cite{Angioni2006}. In this fluid description, the diffusivity depends 
upon $Z$ directly and indirectly due to the fluid curvature. For impurities in the trace limit, the turbulent convective velocity can be decomposed as \cite{Angioni2009}
\begin{equation}\label{turbulent_convective velocity}
V^{turb}_I=\frac{\chi_{e}^{eff}}{\chi_{e}^{\textsc{gkw}}}\frac{D_I^{\textsc{gkw}}}{a}\left(C_T\frac{a}{L_{T_i}}+C_uu'+C_p\right)
\end{equation}
where $C_T$, $C_u$ and $C_p$ are the thermodiffusion, rotodiffusion and convective dimensionless turbulent transport coefficients respectively \cite{Angioni2006}. $V^{turb}$ 
moderately increases with $Z$ because $C_u$ is dependent on the impurity Mach number; the ratio of the two impurity Mach numbers is approximately \mbox{$M_{C}/M_{He}\approx\sqrt{m_{C}/m_{He}}\sim1.7$}. 
We also noted that $C_T$ was moderately smaller for carbon compared to helium, however this decrease was modest compared to the increase in $C_u$. There is little change in $C_p$ since this 
depends upon the ratio $Z/A$, which is constant for fully ionised impurities.   

To determine the uncertainties in the modelled transport coefficients, the simulations have been repeated whilst systematically varying each 
experimental input according to their standard deviation. The biggest uncertainty in $D_I^{\textsc{neo}}$ was caused by the $q$ profile ($\Delta q/q\sim15$ \%), which is 
unsurprising considering that both low and high collisionality neoclassical diffusivities are proportional to $q^2$. $V_I^{\textsc{neo}}$ was equally sensitive to the error in the temperature and 
density radial gradient ($\sim40$ \%). Bremsstrahlung measurements give $Z_{eff}$ profiles that are typically higher than from CX measurements in similar L-mode plasmas, with 
$Z_{eff}$ increasing up to $\sim2$ in the region $\rho>0.5$ \cite{Patel2004}. We explored this uncertainty by increasing the CX C$^{6+}$ concentration by a factor of 10 in the
plasma edge and redoing the \textsc{neo} simulations, but the corresponding errors were modest compared to those arising from uncertainties in $q$ and in the density and temperature gradients.
The main source of uncertainty in the gyrokinetic diffusivity arises from the error in $\chi_{e}^{eff}$ ($\sim30$ \%), while the convective velocity uncertainty is sensitive to both $\chi_{e}^{eff}$ and the ion temperature gradient (which multiplies 
the thermodiffusion coefficient). 

Modelled transport coefficients for helium and carbon at $\rho=[0.6,0.7]$, including both neoclassical and turbulent contributions, are shown by the square and circle symbols respectively in 
figure \ref{fig:3.4} and summarised in table \ref{table:transport}. Note that symbols are also shown (for carbon) within $\rho\leq0.5$ to indicate the typical error bar of the 
neoclassical impurity transport coefficients. The magnitude of the carbon and helium transport coefficients has been 
reproduced within the uncertainties for $\rho=0.6$, however the magnitude of the experimental convective velocity for carbon is significantly higher than the model prediction at $\rho=0.7$. 
Encouragingly, the modelled trend in $Z$ for both diffusion and convective velocity rates, given by $(D_C^{\textsc{neo}}+D_C^{\textsc{gkw}})/(D_{He}^{\textsc{neo}}+D_{He}^{\textsc{gkw}})$ and similarly for the
convective velocity, agrees within the experimental uncertainty of $D_C/D_{He}$ and $V_C/V_{He}$ at both $\rho=0.6$ and $\rho=0.7$. Note that the uncertainty in the ratio of the modelled diffusivity for carbon 
and helium (or in the corresponding ratio for the modelled convective velocities) is small because of error cancellation.

Most noticeable from the experimental impurity transport coefficients shown in figure \ref{fig:3.4}, is that the dependence on $Z$ agrees with the trends expected for 
TEM turbulence that is predicted to be unstable. This is opposite to the trend expected for ITG turbulence, where turbulent diffusion is expected to decrease with $Z$, while 
both thermodiffusion and rotodiffusion are directed outwards \cite{Angioni2006,Casson2010}. Thus, the main conclusion from this work is that a combination of neoclassical and TEM driven 
transport is most likely causing the observed impurity particle flux in the outer region of the plasma. Furthermore the model, combining neoclassical transport with quasi-linear turbulence,
is shown to provide reasonable estimates of the impurity transport coefficients over the analysed spatial region.
 
\section{Conclusion}\label{sec:5}

This paper has evaluated fully ionised carbon and nitrogen impurity transport coefficients during two repeat L-mode 900 kA MAST discharges for comparison with earlier results for helium.
There is little difference between the carbon (intrinsic) and nitrogen (extrinsic) transport coefficients which is not surprising, considering that both species are fully ionised over the majority of 
the plasma radius and therefore have a similar atomic charge and mass. Both experience a diffusivity of the order of \mbox{1-10 m$^2$/s} and 
a strong inward convective velocity of \mbox{$\sim40$ m/s} near the plasma edge, and a region of outward convective velocity at mid-radius. In comparison with helium, carbon has higher
diffusivity and convective velocity near the plasma edge and furthermore its convective velocity has a reversal at mid-radius. 

Neoclassical and quasi-linear gyrokinetic simulations 
of the particle flux have been run with \textsc{neo} and \textsc{gkw} respectively. Neoclassical transport alone is sufficient to explain the observed impurity 
transport of each species within $\rho\leq0.4$, but cannot explain the magnitudes of the transport coefficients or trend in $Z$ in the region $0.4<\rho\leq0.8$. The reversal of the convective velocity direction 
at mid-radius is consistent with neoclassical Banana-Plateau transport. A quasi-linear approach has been used to calculate the magnitude of turbulent diffusivity 
and convective velocity, and indicates that in these MAST discharges the dominant linear mode is in the trapped electron mode wavenumber region (because of the suppression of ITG turbulence by sheared toroidal flow).
Summing the neoclassical and quasi-linear turbulent contributions to the particle flux provides a model of the impurity transport that reasonably describes its magnitude and dependence on 
charge; however it is noted that the observed carbon transport at $\rho=0.7$ is still significantly higher than the model. For helium, the anomalous convective velocity is dominated by 
thermodiffusion, whereas both rotodiffusion and thermodiffusion contribute to the anomalous carbon convective velocity. In summary, this paper provides further support for the existence of 
trapped electron modes in MAST plasmas and their impact on low-$Z$ impurity transport. 

\section*{Acknowledgements}
SSH would like to thank Neil Conway for his diagnostic help. This work has been part-funded by the RCUK Energy Programme [grant number EP/I501045]. To obtain further information on the data and models underlying this paper please contact PublicationsManager@ccfe.ac.uk. The views and opinions expressed herein do not necessarily 
reflect those of the European Commission. The gyrokinetic calculations were carried out on the ARCHER supercomputer using resources allocated to the Plasma HEC Consortium (EPSRC grant EP/L000237/1).

\section*{References}
\bibliographystyle{prsty}
\bibliography{myreferences}

\begin{thebibliography}{10}

\bibitem{Kallenback2013}
A. Kallenbach {\it et~al.}, Plasma Phys. Control. Fusion\/ {\bf 55},  124041
  (2013).

\bibitem{Guirlet2006}
R. Guirlet {\it et~al.}, Plasma Phys. Control. Fusion\/ {\bf 48},  B63  (2006).

\bibitem{Angioni2006}
C. Angioni and A.~G. Peeters, Phys. Rev. Lett. {\bf 96},  095003  (2006).

\bibitem{Wenzel1990}
K.~W. Wenzel and D.~J. Sigmar, Nucl. Fusion {\bf 30},  1117  (1990).

\bibitem{Wong1987}
K.~L. Wong and C.~Z. Cheng, Phys. Rev. Lett. {\bf 59},  2643  (1987).

\bibitem{Synakowski1990}
E.~J. Synakowski {\it et~al.}, Phys. Rev. Lett. {\bf 65},  2255  (1990).

\bibitem{Synakowski1993}
E.~J. Synakowski {\it et~al.}, Phys. Fluids B {\bf 5},  2215  (1993).

\bibitem{Wade1995}
M. Wade {\it et~al.}, Phys. Plasmas\/ {\bf 2},  2357  (1995).

\bibitem{Pasini1990}
D. Pasini {\it et~al.}, Nucl. Fusion {\bf 30},  2049  (1990).

\bibitem{Dux1999}
R. Dux {\it et~al.}, Nucl. Fusion {\bf 39},  1509  (1999).

\bibitem{Giroud2007}
C. Giroud {\it et~al.}, Nucl. Fusion {\bf 47},  313  (2007).

\bibitem{Dux2004}
R. Dux {\it et~al.}, 20th IAEA Fusion Energy Conference, Vilamoura, Portugal,
  CD-ROM file EX/P6-14  (2004).

\bibitem{Valisa2011}
M. Valisa {\it et~al.}, Nucl. Fusion {\bf 51},  033002  (2011).

\bibitem{Casson2015}
F.~J. Casson {\it et~al.}, Plasma Phys. Control. Fusion\/ {\bf 57},  014031
  (2015).

\bibitem{Stutman2003}
D. Stutman {\it et~al.}, Phys. Plasmas\/ {\bf 10},  4387  (2003).

\bibitem{Delgado2011}
L. Delgado-Aparicio {\it et~al.}, Nucl. Fusion {\bf 51},  083047  (2011).

\bibitem{Scotti2013}
F. Scotti {\it et~al.}, Nucl. Fusion {\bf 53},  083001  (2013).

\bibitem{Henderson2014}
S.~S. Henderson {\it et~al.}, Nucl. Fusion {\bf 54},  093013  (2014).

\bibitem{Roach2005}
C.~M. Roach {\it et~al.}, Plasma Phys. Control. Fusion\/ {\bf 47},  B323
  (2005).

\bibitem{Lauro-Taroni1994}
L. Lauro-Taroni {\it et~al.}, 21st EPS Conference on Controlled Fusion and
  Plasma Physics, Montpellier, France {\bf 18B},  102  (1994).

\bibitem{Belli2008}
E.~A. Belli and J. Candy, Plasma Phys. Control. Fusion\/ {\bf 50},  095010
  (2008).

\bibitem{Belli2012}
E.~A. Belli and J. Candy, Plasma Phys. Control. Fusion\/ {\bf 54},  015015
  (2012).

\bibitem{Peeters2009}
A.~G. Peeters {\it et~al.}, Comput. Phys. Commun. {\bf 180},  2650  (2009).

\bibitem{HendersonThesis}
S.~S. Henderson, Ph.D. thesis, University of Strathclyde, 2014.

\bibitem{Scannell2008}
R. Scannell {\it et~al.}, Rev. Sci. Instrum. {\bf 79},  10E730  (2008).

\bibitem{Conway2006}
N.~J. Conway {\it et~al.}, Rev. Sci. Instrum. {\bf 77},  10F131  (2006).

\bibitem{Patel2004}
A. Patel, P. Carolan, N. Conway, and R. Akers, Rev. Sci. Instrum. {\bf 75},
  4944  (2004).

\bibitem{Appel2006}
L. Appel {\it et~al.}, 33rd EPS Conference on Plasma Physics, Rome, Italy {\bf
  30I},    (2006).

\bibitem{DeBock2008}
M.~F.~M. {De Bock} {\it et~al.}, Rev. Sci. Instrum. {\bf 79},  10F524  (2008).

\bibitem{Summers2004}
H.~P. Summers, The ADAS User Manual {\bf V2.6},  http://adas.phys.strath.ac.uk
  (2004).

\bibitem{Wesson1997}
J.~A. Wesson, Nucl. Fusion {\bf 37},  577  (1997).

\bibitem{Tamor1981}
S. Tamor, J. Comput. Phys. {\bf 40},  104  (1981).

\bibitem{Estrada-Mila2005}
C. Estrada-Mila, J. Candy, and R. Waltz,  {\bf 12},  022305  (2005).

\bibitem{Kotschenreuther1995}
M. Kotschenreuther, G. Rewoldt, and W. Tang, Comput. Phys. Commun. {\bf 88},
  128  (1995).

\bibitem{Dorland2000}
W. Dorland, F. Jenko, M. Kotschenreuther, and B. Rogers, Phys. Rev. Lett. {\bf
  85},  5579  (2000).

\bibitem{Roach2009}
C.~M. Roach {\it et~al.}, Plasma Phys. Control. Fusion\/ {\bf 51},  124020
  (2009).

\bibitem{Cenacchi1988}
G. Cenacchi and A. Taroni, Report JET-IR {\bf 88},  03  (1988).

\bibitem{Cenacchi1988a}
G. Cenacchi and A. Taroni, Rapporto ENEA RT/TIB {\bf 88},  5  (1988).

\bibitem{Angioni2011}
C. Angioni {\it et~al.}, Nucl. Fusion {\bf 51},  023006  (2011).

\bibitem{Casson2013}
F.~J. Casson {\it et~al.}, Nucl. Fusion {\bf 53},  063026  (2013).

\bibitem{Angioni2014a}
C. Angioni {\it et~al.}, Nucl. Fusion {\bf 54},  083028  (2014).

\bibitem{Angioni2009}
C. Angioni {\it et~al.}, Nucl. Fusion {\bf 49},  055013  (2009).

\bibitem{Casson2010}
F.~J. Casson {\it et~al.}, Phys. Plasmas\/ {\bf 17},  102305  (2010).

\end{thebibliography}
\end{document}